\documentclass[pdflatex,sn-mathphys-num]{sn-jnl}

\usepackage{graphicx}%
\usepackage{multirow}%
\usepackage{amsmath,amssymb,amsfonts}%
\usepackage{amsthm}%
\usepackage{mathrsfs}%
\usepackage[title]{appendix}%
\usepackage{xcolor}%
\usepackage{textcomp}%
\usepackage{manyfoot}%
\usepackage{booktabs}%
\usepackage{algorithm}%
\usepackage{algorithmicx}%
\usepackage{algpseudocode}%
\usepackage{listings}%

\theoremstyle{thmstyleone}%
\theoremstyle{thmstyletwo}%

\theoremstyle{thmstylethree}%

\begin{document}

\title[Article Title]{Millimetre-Wave Comb Generated by an Optical Microcomb}








\author*[1,2]{\fnm{L.} \sur{Peters}}\email{l.peters@lboro.ac.uk}
\author[1,2]{\fnm{A.} \sur{Cutrona}}
\author[1]{\fnm{A. R.} \sur{Cooper}}
\author[1,2]{\fnm{L.} \sur{Olivieri}}
\author[1]{\fnm{F.} \sur{Getman}}
\author[1,2]{\fnm{V.} \sur{Cecconi}}
\author[1]{\fnm{N.} \sur{Paul}}
\author[1,2]{\fnm{D.} \sur{Das}}
\author[2]{\fnm{M.} \sur{Rowley}}
\author[3]{\fnm{S. T.} \sur{Chu}}
\author[4]{\fnm{B. E.} \sur{Little}}
\author[5]{\fnm{R.} \sur{Morandotti}}
\author[6]{\fnm{D. J.} \sur{Moss}}
\author[1,2]{\fnm{J. S. Totero} \sur{Gongora}}
\author[1,2]{\fnm{A.} \sur{Pasquazi}}
\author*[1,2]{\fnm{M.} \sur{Peccianti}}\email{m.peccianti@lboro.ac.uk}

\affil[1]{\orgdiv{Emergent Photonics Research Centre (EPicX), Dept. of Physics}, \orgname{Loughborough University}, \orgaddress{\city{Loughborough}, \postcode{LE11 3TU}, \country{UK}}}

\affil[2]{\orgdiv{Emergent Photonics Lab (EPic), Department of Physics and Astronomy}, \orgname{University of Sussex}, \orgaddress{\city{Brighton}, \postcode{BN1 9QH}, \country{UK}}}

\affil[3]{\orgdiv{Department of Physics}, \orgname{City University of Hong Kong}, \orgaddress{\street{Tat Chee Avenue}, \city{Hong Kong}, \country{China SAR}}}

\affil[4]{\orgname{QXP Technologies, Inc.}, \orgaddress{\city{Xi’an}, \postcode{710311}, \country{China}}}

\affil[5]{\orgname{INRS-EMT}, \orgaddress{\street{1650 Boulevard Lionel-Boulet}, \city{Varennes}, \state{Québec}, \postcode{J3X 1S2}, \country{Canada}}}

\affil[6]{\orgdiv{Optical Sciences Centre}, \orgname{Swinburne University of Technology and ARC-COMBS}, \orgaddress{\city{Hawthorn}, \state{VIC}, \postcode{3122}, \country{Australia}}}


\abstract{Metrological-grade millimetre wave baseband comb sources covering the sub-terahertz window are a key building block for next-generation wireless communications, precision sensing, and positioning systems. While optical microcombs have set new benchmarks in ultra-low phase noise single-frequency microwave generation, to date, no microcomb source has directly produced a millimetre-wave baseband comb. Here, we present a 50 GHz repetition rate carrier-envelope offset stabilised millimetre-wave baseband comb source covering the sub-terahertz region, generated from an optical microcomb source. Our microresonator-filtered microcomb enables direct, coherent downconversion via photoconductive antennas, even without external amplification. The metrological-grade optical soliton source produces single-cycle, naturally zero carrier-envelope offset millimetre wave baseband combs. It supports time-domain spectroscopy without any need to temporally align the source and detection pulses, as the ultra-high phase coherence allows significant differences between the optical paths of the source and detection pulses, which we tested over 8m, finding no degradation even in free-running operation. Finally, the multisoliton operation regime provides a simple way of spectrally tailoring the microwave output by selecting different optical soliton states.}




\maketitle

\section{Introduction}\label{sec1}

The development of metrological grade millimetre and terahertz (THz) baseband wave sources is critical for applications in communication \cite{Kang2024Frequency,Wang2024On,Leitenstorfer20232023,You2020Towards,Ma2018Security},  sensing \cite{Leitenstorfer20232023,Jackson2011Survey,Federici2005THz,EguchiLamb,Djevahirdjian2023Frequency,Longhai2017Finger,Zhong2019Progress}, timing and positioning \cite{Ghelfi2014fully,Yang2022Frequency,Zhao2024All,Wu2023Vernier,Kudelin2024Photonic,Sun2024Integrated,Wildi2024Phase,Jin2025Microresonator}.  The next generation of millimetre wave (mm-wave) frequency standards in the sub-THz range will form the foundation of future 6G communication systems and positioning devices \cite{You2020Towards,Ma2018Security,Yang2022Frequency}. It will require precision time-traceability \cite{Kang2019Free,Shin2023Photonic} and low phase noise \cite{Djevahirdjian2023Frequency}. 
Over the past two decades, mm-wave sources have become progressively linked to photonics and, in particular, to optical frequency combs \cite{Udem2002Optical,Picque2019Frequency,Diddams2020Optical}, which  presently enable the most advanced optical metrological systems, both in the spectral and time domains \cite{Udem2002Optical,Picque2019Frequency,Diddams2020Optical}. This includes metrological-grade optical frequency combs that can generate ultralow-noise single frequencies in the mm-wave baseband regime. Prominent related examples  are quantum cascade lasers \cite{Barbieri2010Phase, Ravaro2011Phase,Faist2016Quantum,Freeman2017Injection,Consolino2019Fully,Guerrero2024Harmonic} and electronic oscillators \cite{Hiraoka2022Passive,Arikawa2024Phase,Ummethala2019THz}, which dramatically improve spectral purity and frequency accuracy when stabilised using optical frequency combs. These hybrid systems are particularly important for spectroscopy applications, including the interrogation of quantum states \cite{EguchiLamb,Zhang2024Floquet}, time-domain THz spectroscopy \cite{Cherniak2023Laser}, and dual-comb schemes \cite{Hsieh2014Spectrally,Jia2022Integrated,Li2023Terahertz}. More broadly, optical frequency combs  that can be utilised to generate baseband mm-wave and signals are expected to be a critical building block of next-generation wireless communications and positioning systems, both as communication sources, e.g. as multichannel solutions for  parallel frequency-division multiplexing and multi-user links or as calibration and metrological devices \cite{Arikawa2024Phase,Jia2022Integrated,Ma2017Frequency}.  Compatibility with coherent optical links \cite{Yang2022Frequency,Kang2019Free}, especially for clock-stabilised frequency distribution, is a key feature.
In this scenario, microcomb-based mm-wave sources \cite{Kudelin2024Photonic,Sun2024Integrated,Zhao2024All} have demonstrated compactness and unprecedented spectral purity, especially when the microresonator combs \cite{Marin2017Microresonator,Brasch2016Photonic,Huang2016broadband,Spencer2018optical,Hu2018Single,Fulop2018High,Xu202111,Liu2020Monolithic,Meng2020Mid,Boggio2022Efficient,Wu2025Vernier,Kudelin2024Photonic,Sun2024Integrated,Zhao2024All} are operated in the so-called soliton regime \cite{Kivshar2003Optical,Lugiato2015Nonlinear,2016Nonlinear}. Because microcombs are based on monolithic, integrated resonators, they have intrinsically superior phase noise performance, particularly in their repetition rate, which naturally falls in the sub-THz range. This specific  combination of features makes microcomb-driven mm-wave generation highly desirable compared to solutions based on traditional mode-locked pulsed lasers \cite{Jeong2020Ultralow,Kwon2022Ultrastable,Bao2021Quantum,Matsko2013On}.
Recent demonstrations \cite{Zhang2019Terahertz,Tetsumoto2021Optically} achieved ultra-low phase noise in the microcomb repetition rate that was efficiently transferred to the mm-wave domains, enabling the creation of ultra-high-purity carriers. Here, the excellent phase noise achieved in the optical regime was moved to lower frequencies via optical division \cite{Wu2023Vernier,Kudelin2024Photonic,Sun2024Integrated,Wildi2024Phase,Jin2025Microresonator},  producing state-of-the-art high-purity, single-frequency microwave sources \cite{Shin2023Photonic,Zhang2019Terahertz,Tokizane2023Terahertz,Tokizane2024Wireless,Yao2022Soliton,Kuse2022Low,Kuse2022Amplification}.  

Beyond single frequency generation, the next major frontier is the direct generation of a broadband coherent frequency comb in the sub-THz regime, important for channel-multiplexed communications, broadband spectroscopy, and the precision evaluation of emerging 6G devices \cite{Kang2024Frequency,Wang2024On,Leitenstorfer20232023,You2020Towards,Ma2018Security}.

Here, we demonstrate a frequency comb directly generated in the sub-THz baseband regime by an optical microcomb source, using photoconductive antennas for optical-to-terahertz conversion. This simple, yet practical approach achieves carrier envelope offset-free mm-wave frequency combs, which we then employ to demonstrate a fully microcomb-driven THz time-domain spectroscopy (TDS) system. Such a methodology differs from the route used so far for mm-wave generation via microcombs \cite{Zhang2019Terahertz,Lucas2020Ultralow,Kwon2022Ultrastable,Sun2024Integrated,Kudelin2024Photonic,Zhao2024All}, which has relied on photodiode technologies \cite{Tokizane2023Terahertz,Jia2022Integrated}. THz-TDS, in particular, requires intrinsically long-term robustness, stability, and high spectral efficiency. These features are generally challenging for most microcomb implementations,  especially for externally driven schemes that are inherently more sensitive to environmental fluctuations. However, they are well exhibited by laser cavity-solitons in a nested-cavity microcomb laser \cite{Bao2019Laser,Rowley2022Self}, together with the low jitter and long-delay coherence we demonstrate here, our microresonator-filtered fibre laser (Fig. 1a) operating via a laser cavity-soliton (Fig. 1b), producing a 50 GHz, intrinsically carrier-envelope offset-free millimetre-wave baseband comb source covering the sub-THz region (Fig. 1c–e). The laser cavity-soliton microcomb features very high spectral efficiency, enabling direct interfacing with optical-to-THz nonlinear conversion and detection schemes. Importantly, we can drive the full time-domain spectroscopy without any additional amplification stage (Fig. 1c).  Our technology is compatible with metrological-grade microcombs and, remarkably, the spectroscopy conversion remains coherent after long spatial delays (above 8 metres) between the source and the detection pulses, even with the free-running source. Beyond demonstrating single-soliton direct conversion, we further exploit the intrinsic ability of microcombs to host multiple solitons, showing that their mutual delay provides a natural mechanism to shape the mm-wave comb spectrum (Fig. 1d–e). This relies on the simple principle that pulses with different delays interfere, naturally redistributing the spectral power in the generated mm-wave comb. In multisoliton microcombs, such spectral shaping arises directly from the source configuration, without requiring additional components. Together, these results demonstrate the direct mm-wave baseband comb from a microcomb, combining single-soliton robustness with multi-soliton spectral shaping, and opening perspectives for sub-THz communications, spectroscopy, and precision metrology.

\section{Results}\label{sec2}
A schematic of the experimental setup is presented in Fig. 1a. The microcomb laser \cite{Bao2019Laser,Rowley2022Self} is based on a nonlinear microresonator closed in an erbium-doped fiber amplifier (EDFA) cavity. A typical laser cavity-soliton microcomb produced by the system is shown in Fig. 1b, with a 48.91 ($/sim$ 50) GHz repetition rate. The millimeter-comb source/detection system is based on a commercially available THz-TDS setup components. It is fed directly by laser cavity-solitons produced by the microcomb laser at the output (see Methods for details) of the resonator. The TDS comprises two commercial photoconductive antennas fed by two output replicas delayed by a mechanical optical delay line. Specifically, a photoconductive antenna generates a millimeter waveform $E_{THz} (t)$ that follows the derivative of the intensity optical profile $I(t)$  feeding it \cite{Matsuura1997Generation}:

\begin{equation}
E_{THz}{(t)} \propto \frac{\delta{I(t)}}{\delta{t}}. \label{eq1}
\end{equation}

The receiver antenna is configured to perform photoconductive sampling of the emitted field transient, reconstructing the full temporal waveform of the impinging broadband field, as opposed to simply extracting the narrowband notes typical of heterodyne mm-wave detectors. Further details regarding the emission mechanisms and the setup specifications are provided in the Methods section, but generally, the operation implemented by the TDS reconstruction is the following convolution:
\begin{equation}
E_{THz}^{det}(t) = E_{THz}(t)* I(t). \label{eq2}
\end{equation}
Figure 1c shows the conversion of a free-running, single-soliton case, with about 5 mW output power, used to feed the two photoconductive switches without any optical post-processing. The free-running microcomb laser consumes only the electrical power necessary to run the 980 nm EDFA pump, which is below 1.5 W in this configuration. Although the measurement required several seconds of averaging per data point (see Methods for full details), this demonstrates the main compatibility feature of our platform with commercial photoconductive THz systems. The pulse profile is remarkably single-cycle, well reproducing a differential pulse as expected by Eq. (1), with a peak-to-peak duration (half-cycle) on the order of 1.5–2 ps. Such a duration is consistent with the general output pulse width of the comb (roughly 500 fs). It is important to stress how, in our realisation, the limits imposed by off-the-shelf solutions play out. Indeed, commercial THz photoconductive switches (as the ones we utilised) are designed as resonant antennas for standard pulsed lasers with repetition rates ($\sim$100 MHz) much lower than the $\sim$ 50 GHz of our microcomb source and exhibit reduced conversion efficiencies outside the sub-THz region, highlighting a tremendous potential for improvement in efficiency.
To provide comprehensive diagnostics and overcome such limitations, we filtered and amplified the comb to increase the signal-to-noise ratio (SNR); see Methods for details. Examples of power spectra extracted for single (Fig. 1d) and two (Fig. 1e) soliton cases show the comb feature with $\sim$ 50 GHz spacing of the generated mm-wave, which we discuss in greater detail in Figs. 2 and 3.
\begin{figure}[h]
\centering
\includegraphics[width=0.9\textwidth]{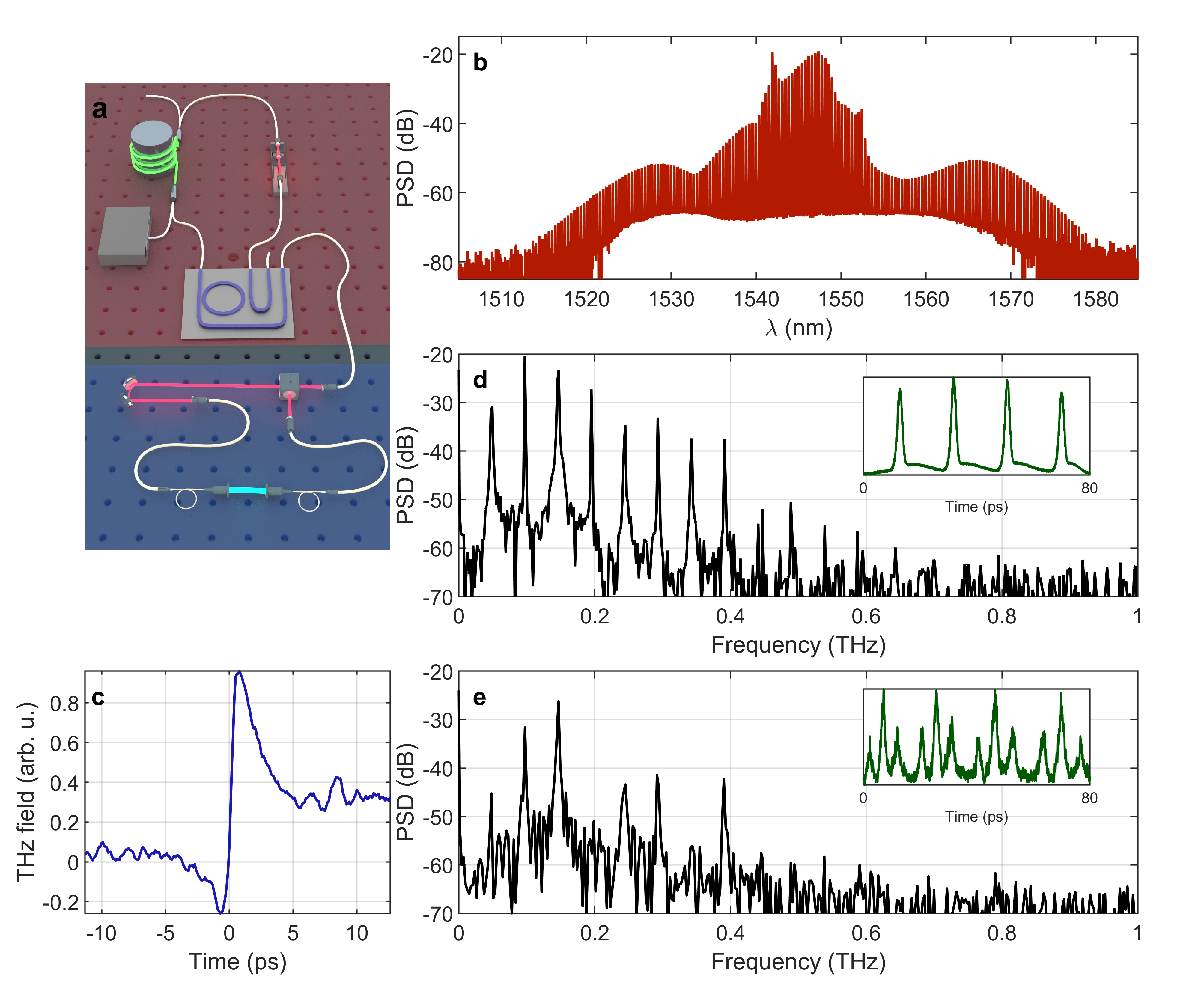}
\caption{\textbf{Generation of a millimetre-wave comb via photoconductive conversion of a laser cavity-soliton microcomb.}
\textbf{a} Experimental setup. A microcomb laser, consisting of a nonlinear microresonator (FSR 48.9 ($/sim$ 50) GHz) filtering a fibre laser cavity loop (FSR $\sim$ 95MHz) (red region), is coupled to a time-domain spectroscopy (TDS) system (blue region) via a beam splitter feeding both transmitter and receiver photoconductive-switch antennas. A mechanical stage controls the relative pulse delay between the pulse replicas. \textbf{b}  A typical optical spectrum of a single-soliton microcomb state.
\textbf{c} TDS trace of a single-cycle millimetre-wave pulse obtained by feeding the TDS setup with a single-soliton microcomb (roughly mW average power), without further processing or amplification. The slight pedestal variation in the time-domain trace is attributed to system drift during mechanical scanning.
\textbf{d} Power spectrum (PSD) of the millimetre-wave trace from c, highlighting the comb structure; the inset shows the autocorrelation. \textbf{e} Same as d, but for a state composed of two non-equidistant solitons, showing a distinctly different millimetre-wave trace.}\label{fig1}
\end{figure}

Figure 2 provides an overview of the system performance in the single-soliton regime. Figure 2a shows the long-term behaviour of the microcomb, displaying the optical spectrum recorded over roughly one hour. The TDS field trace was reconstructed over a temporal window exceeding 500 ps, covering approximately 25 pulses at the microcomb repetition rate of $\sim$ 50 GHz. Notably, although TDS scanning is a typically slow process, the LCS microcomb demonstrated high stability throughout the scan (around 6000 samples with a second-scale lock-in amplifier integration time), reflecting its inherent self-recovery nature\cite{Bao2019Laser,Rowley2022Self}. Figures 2a and 2b both illustrate the consistent spectral properties tracked during the scan, with the TDS trace (Fig. 2b) exhibiting only long-term DC drifts. The pulse field profile, zoomed in Fig. 2c, is typical for photoconductive-based TDS \cite{Matsuura1997Generation} and resembles a pulse derivative (Ricker wavelet). The calculated power spectral density reveals a bandwidth exceeding 600 GHz within our detection dynamics, with ten resolvable harmonics of the $\sim$ 50 GHz carrier frequency and good coverage of the sub-THz region. The waveform, hence, lies within the typical microwave measurable range and is directly obtained from the rectification of a coherent optical pulse, with all the harmonics inherently coherent with a zero carrier-envelope offset.
\begin{figure}[h]
\centering
\includegraphics[width=0.9\textwidth]{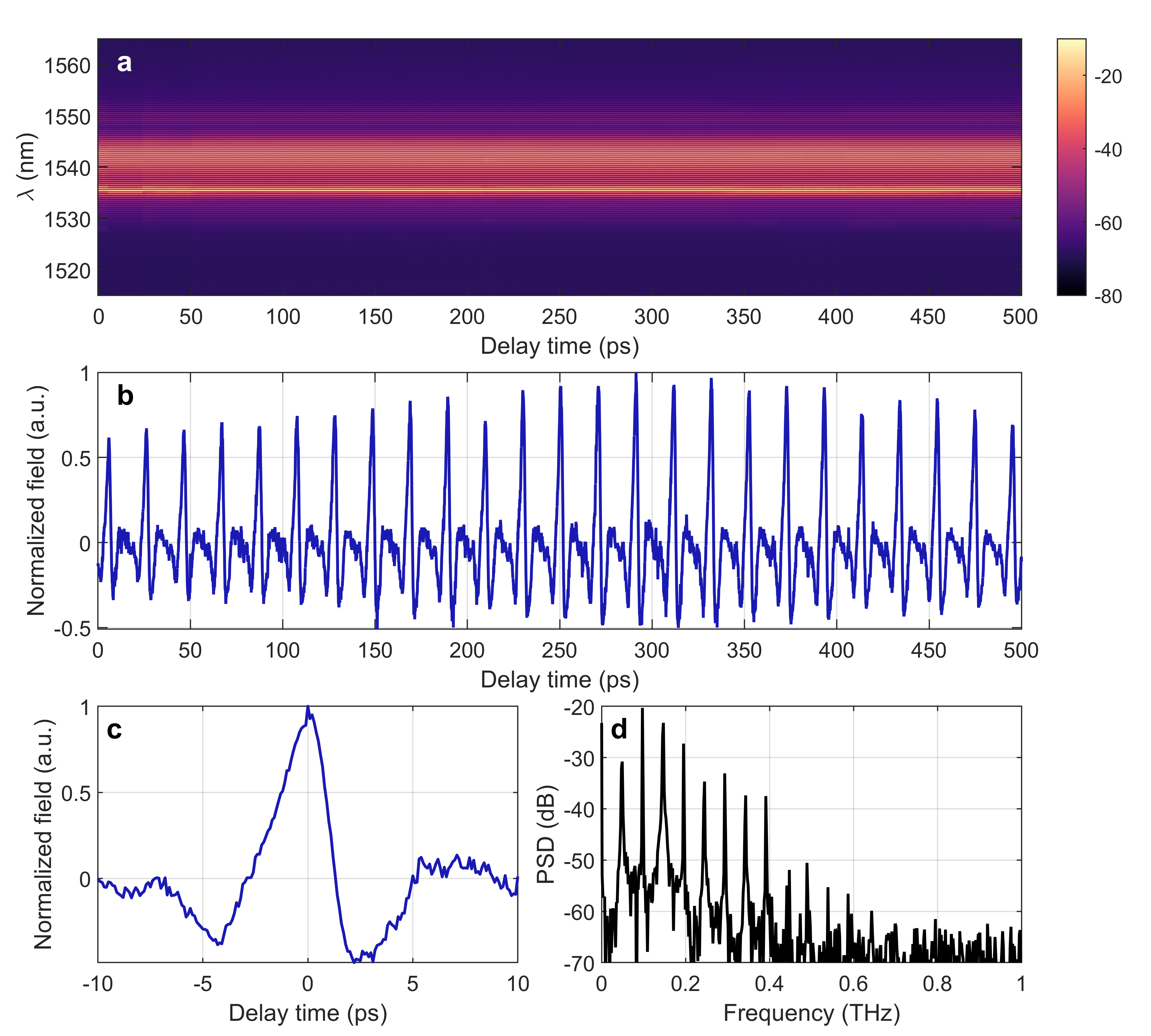}
\caption{\textbf{Long term robustness of THz-wave generation for single soliton states}
\textbf{a} False-colour map in dB on the colour axis of the optical spectrum analyser traces during the TDS-scan, showing wavelength vs time delay step. \textbf{b}  TDS-trace obtained converting the microcomb pulse extracted at $\sim$ time delay 290 ps.  \textbf{c}.  single cycle trace extracted from b  \textbf{d}. Power spectral density of the waveform b.}\label{fig2}
\end{figure}

We then explored the system’s capability to control and tune the millimetre-wave comb emission. Multisoliton states generally manifest as non-equidistant series of pulses, and spectral shaping and control based on soliton crystals is well-studied in the microcomb literature \cite{Cole2017Soliton,Zhang2020Spectral}. The simplest scenario is that of a two-soliton state, where the spectrum depends directly on the relative delay between the two soliton pulses. The results shown in Fig. 1e demonstrate a markedly different spectral distribution compared to the single-soliton state.
Figure 3 offers a more detailed analysis of a two-soliton state, visible in the autocorrelation traces (Fig. 3a for the temporal evolution, Fig. 3b for a single trace, and Fig. 3c for the reconstructed intensity profile), with a relative delay of 6.4 ps—approximately $32\%$ of the period. Multisoliton states tend to pin to defects, preserving the configuration if not strongly perturbed. The state remained stable throughout the 30-minute scan (as seen in both the autocorrelation evolution, Fig. 3a, and the OSA traces, Fig. 3d, producing a uniform millimetre waveform in Fig. 3e.
Examining a single period in Fig. 3f, we observe three pulses, with a central dominant peak. The numerical waveform in Fig. 3g, obtained by differentiating the pulse train in Fig. 3c, represents the effective generating field created by the double soliton state. The TDS system, however, retrieves its convolution with the optical pulse train in Fig. 3c, which is shown in Fig. 3h, following Eq. (2), and aligns well with the experimental TDS trace.
The spectral trace in Fig. 3e, obtained via Fourier transformation of the TDS data, shows the characteristic signature of interference. However, the harmonics are somewhat suppressed in the retrieved trace due to the temporal convolution performed by the TDS operation showin in  in Fig. 3f (see Methods for details). Extended Fig. A1 shows a similar case for a different pulse delay, further demonstrating that simple spectral control through pulse delay enables reconfiguration of the spectral balance between the generated harmonics, while Extended Fig. A2 and the methods section offers more detail on the spectral filtering observed in a TDS without equidistant pulses.
\begin{figure}[h]
\centering
\includegraphics[width=0.9\textwidth]{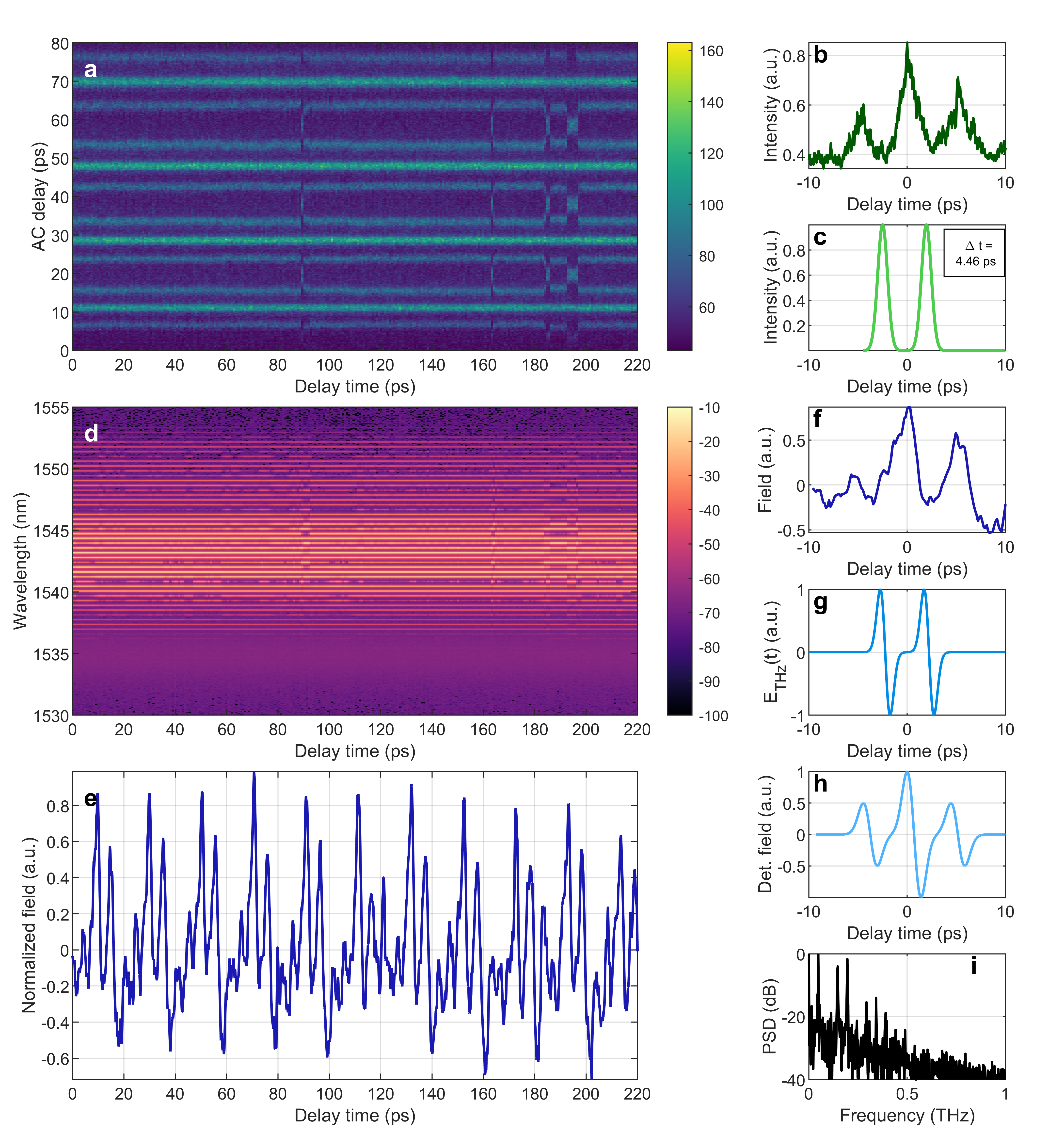}
\caption{\textbf{Long-term robustness and high-signal-to-noise ratio time-domain spectroscopy for a state with two non-equidistant solitons.} \textbf{a} False-colour map in dB on the colour axis of the autocorrelation traces in time. \textbf{b} A typical autocorrelation trace at t $=$ 0. \textbf{c} Reconstructed double soliton state with periodicity 20 ps and soliton spacing 4.46 ps extracted from the autocorrelation trace. \textbf{d} False-colour map of the optical spectrum analyser traces in time
\textbf{e} Experimental terahertz time-domain spectroscopy trace. \textbf{f} Experimental TDS trace for a single period. \textbf{g} Expected generated waveform converting with Eq(1) the optical pulses in c.  \textbf{h}   Reconstructed TDS trace using Eq. (2) with the generated waveform in  g and the optical pulses in c. \textbf{i} Experimental THz spectrum.}\label{fig3}
\end{figure}

In Fig. 4, we illustrate different soliton dynamics of the LCS during a very long TDS scan, where the laser is left under environmental perturbations and independently transitions between states. Interestingly, by positioning at the boundary between a single-soliton region and a multisoliton state (typically linked with high gain and main cavity detuning) in the parameter landscape, we can simply let the system drift with relatively loose temperature stabilisation to observe these transitions. Even under these conditions, the system shows remarkable stability over several tens of minutes, and the transitions are quite distinct and are triggered by substantial environmental fluctuations.
We tracked the autocorrelation evolution during the scan (Fig. 4a), which reveals transitions from fundamental solitons to dual-soliton states and soliton molecules. Fig. 4b shows the reconstructed mm-wave field. Three distinct two-soliton states are identified: Region I is an equally spaced two-soliton state (with a relative delay of 10 ps between pulses), Region II is unequally spaced (with a relative delay of 6.6 ps), and Region III is also unequal (with a relative delay of 5.6 ps). In Region IV, the system transitions to a single-soliton state. The insets in Fig. 4c compare the measured pulse field dynamics (blue plot) with the one estimated from the autocorrelation and a simple bipolar impulse response of the antenna (yellow plot), confirming the interpretation of the field dynamics as the signature of each state. It is evident that the THz emission is affected by the soliton dynamics, with noticeable changes in both phase and spectral content corresponding to each transition.
\begin{figure}[h]
\centering
\includegraphics[width=0.9\textwidth]{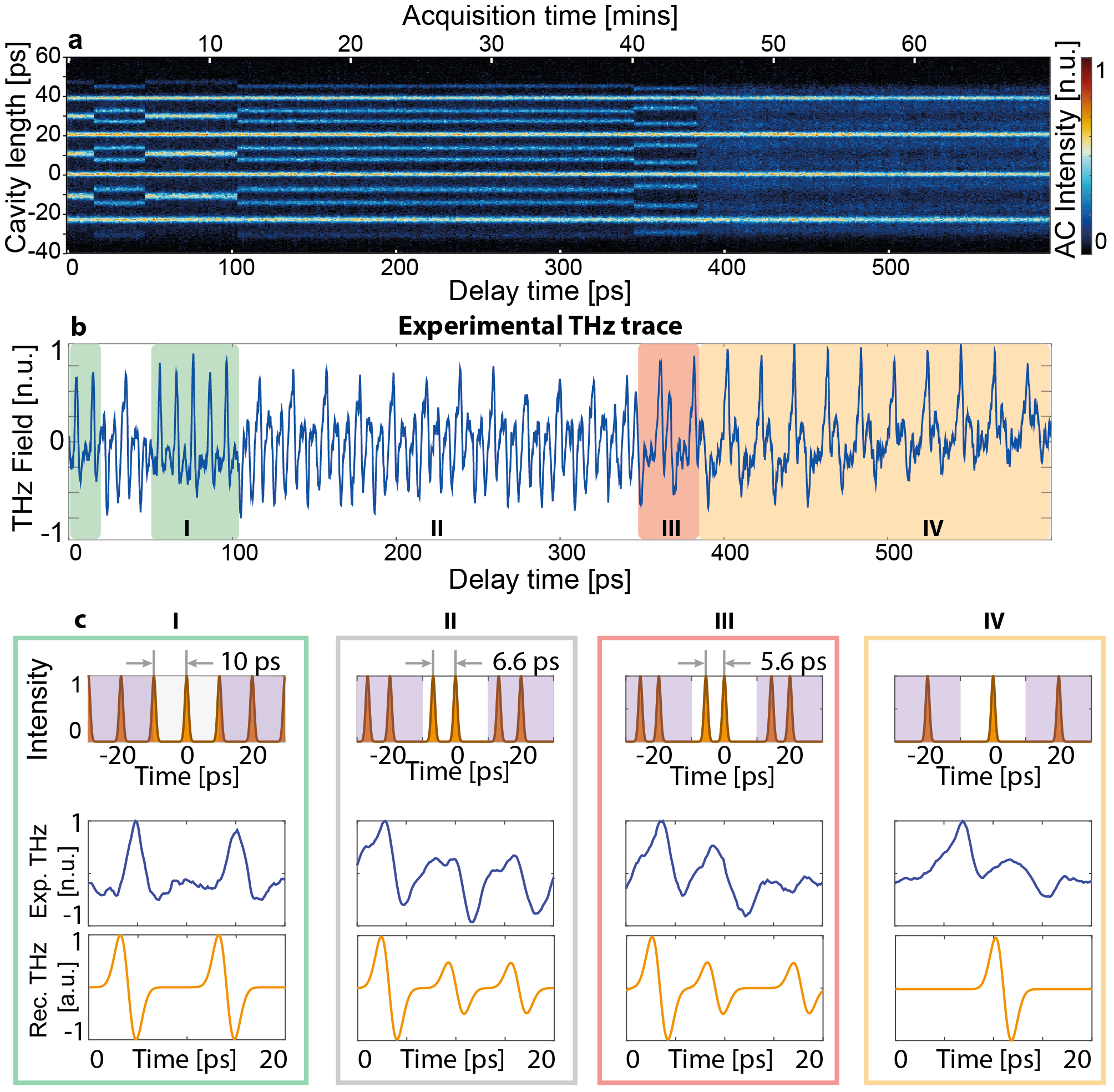}
\caption{\textbf{Terahertz generation from evolving soliton states.} \textbf{a} Experimental autocorrelator evolution map showing transitions between soliton states over the acquisition time. \textbf{b} Experimental terahertz time-domain spectroscopy (TDS) trace, highlighting the distinct THz responses corresponding to different soliton configurations: A – double soliton with equal spacing ($\tau$ = 10 ps, green), B – double soliton with $\tau$ = 6.6 ps (grey), C – double soliton with $\tau$ = 5.6 ps (red), and D – single soliton (orange). \textbf{c} Insets I–IV: For each configuration, the reconstructed optical intensity profile (top), experimental THz trace (middle), and simulated THz trace (bottom) are shown.}\label{fig4}
\end{figure}
Figure 5 explores how stabilising the GHz repetition rate of our laser-cavity-soliton microcomb influences both the optical source and the generated sub-THz comb. The phase noise of the free-running $\sim$ 50 GHz repetition rate is depicted in black in Fig. 5a. In this case, the integrated timing jitter (Fig. 5b), evaluated from 1 MHz down to 10 Hz, is $\sim$ 6.3 ps, but it is strongly affected by a specific low-frequency oscillation at $\sim$ 400 Hz; in the range 600 Hz–1 MHz, the jitter remains well below $\sim$ 150 fs (Fig. 5b). This specific noise contribution can be easily stabilised with mechanical methods, which we applied in our system using a piezo-stretcher on the fibre cavity (see Methods), locking the repetition rate to a microwave GPS-disciplined reference via electro-optic downconversion.

After disciplining the source (phase noise in Fig. 5a, red curve), it strongly improves long-term stability. The high-frequency contribution between 100 kHz and 1 MHz changes only from $\sim$ 79 fs to $\sim$ 75 fs. The lock therefore does not alter the short-term performance. The integrated jitter falls below $\sim$ 150 fs in the full range, showing that the low-frequency portion of the noise is almost completely removed; this is even more visible in the Allan deviation shown in Fig. 5c. It is important to note that, in our measurement, both the repetition rate downconversion scheme and the locking electronics are referenced to the same GPS-disciplined source (see Methods). This partially removes the long-term noise contribution of the reference from the locked measurements, allowing us to evaluate the performance of our system beyond the limitation of our reference source. Within this framework, the Allan deviation falls below $10^{-14}$ above 100 s integration time. This value reflects the intrinsic stability of the microcomb, allowing direct assessment of the system’s long-term performance.

However, this stabilisation has no observable effect on the millimetre-wave comb TDS trace, including output power, spectral shape, or conversion efficiency as show in the example in Fig. 5d. It is important to notice that the high frequency contribution between measurements. Hence within the temporal delay window of our TDS system  (here approximately 500 ps for a one-hour scan), the spectral resolution is not better than a few GHz, which is insensitive to any of the long-term effects seen in Fig. 5a and 5b as a result of stabilisation. The source is naturally  stable enough to support TDS. An important consequence in particular is that TDS can be performed without stringent pump–probe synchronisation, which would be challenging to implement at high repetition rates. Ordinary mechanical delay lines are sufficient even at 50 GHz, and we obtain clean waveforms with optical path differences as large as 8 m between the pulse replicas feeding the generating and detecting antennas—corresponding to about 27 ns and over 1000 replicas. Over such delays, we detect no statistically meaningful deviation in the mm-wave spectrum; any changes in amplitude are attributed to routine coupling-power drift (Fig. 5e and Extended Fig. A3) representing the spectrum of traces in Fig. 5d and 5e. Consequently, the free-running comb remains stable enough for practical spectroscopy over multi-metre delays, while repetition-rate locking—common in optical-frequency-comb metrology—can be applied to this mm-wave platform without compromising its performance. In perspective, since the millimetre-wave comb is created by a purely passive transformation and inherits the stability of the optical comb, this compatibility with metrological sources paves the way for using the mm-wave comb as a multichannel metrological source.

\begin{figure}[h]
\centering
\includegraphics[width=0.9\textwidth]{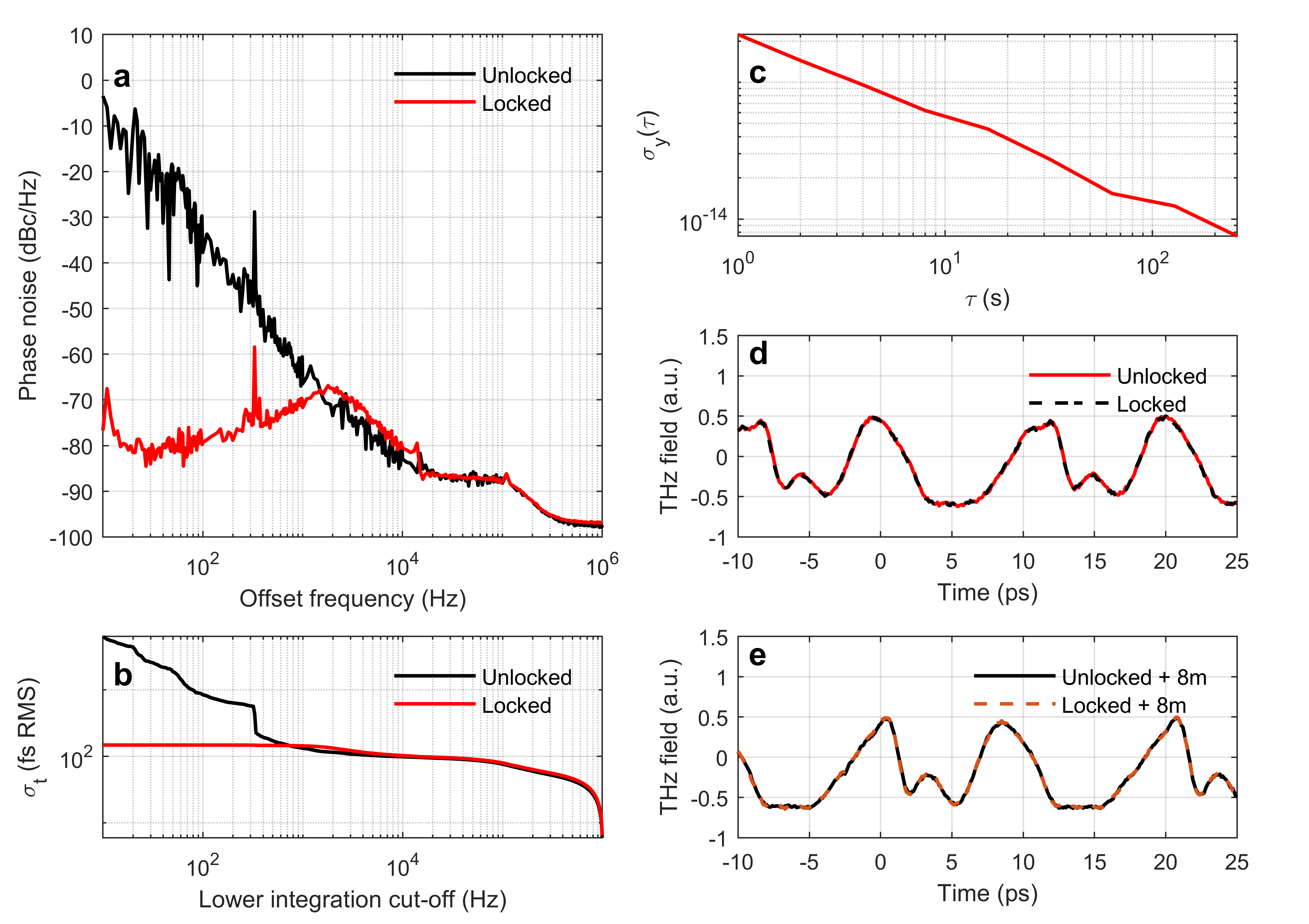}
\caption{\textbf{Phase noise performance of the system. } \textbf{a} Phase noise of an unlocked (black) and repetition rate locked (red) laser cavity-soliton system. \textbf{b} Integrated time jitter for the locked (red) and unlocked (black) cases. \textbf{c} Allan deviation of a repetition rate locked LCS system. \textbf{d} Time‐domain millimetre-wave field comparison of the unlocked (black) and locked (vermillion) cases. \textbf{e} Same as d, but for a TDS measurement where the detection antenna is fed with a pulse replica delayed by 8m (1333rd replica) relative to the generating antenna.}\label{fig5}
\end{figure}

\section{Conclusions}\label{sec4}
We have demonstrated direct millimetre-wave comb generation in the sub-THz range through photoconductive conversion of a laser-cavity-soliton microcomb system. The long-term stability and resilience of the laser to environmental fluctuations enable the use of standard, commercial THz-TDS systems, which—thanks to the high spectral efficiency of our microcomb scheme—can be operated directly without the need for additional amplification, using only modest optical pulse power ($\sim$ 5 mW) and low plug-in electrical power ($<$ 1.5 W). This makes the source highly attractive for practical, portable applications.
The low-jitter performance of the source—even in free-running conditions—allows for relaxed constraints in time-domain spectroscopy, as demonstrated by measurements with delays spanning thousands of pulses. Our results show that microcombs are broadly compatible with nonlinear optical rectification approaches for millimetre-wave generation, unlike established schemes relying on resonant, single-frequency microwave photodiodes. These nonlinear techniques are crucial because they enable single-cycle, inherently carrier-envelope-offset-free comb generation, as clearly demonstrated by our findings. Even when using non-optimised, commercial systems, our approach achieves broadband coverage of the sub-THz range, which is necessary for next-generation wireless communications and positioning systems.
The metrological-grade performance of our source is fully compatible with the photoconductive conversion scheme, confirming the system’s potential as a multichannel metrological source. Finally, the ability to tailor the optical spectrum through multisoliton states—an established feature of microcombs—can be easily leveraged to dynamically shape the power spectral density of the emitted channels.
Looking ahead, further optimisation of the conversion efficiency and integration of the platform with on-chip photonic and antenna technologies present promising prospects for microcomb-driven mm-wave sources, which holds strong potential for metrological-grade applications.

\section{Methods}\label{sec5}
\subsection{Microcomb Setup and Time Domain Spectroscopy Setup}
The experimental system (Extended Fig. A4) relies on a laser cavity-soliton (LCS) generated using a four-port, high-index doped silica integrated resonator with a free spectral range (FSR) of $\sim$ 50 GHz and a linewidth of less than 140 MHz, corresponding to a quality factor of about 1.3 million. This micro-ring resonator was embedded within a polarisation-maintaining ytterbium-erbium co-doped fibre cavity. Specifically, the ring closes the cavity between its input and drop ports, and the through port is used as a laser output. The cavity incorporates a delay line, polarisation optics, an optical isolator, and a tuneable bandpass filter with a 10 nm bandwidth. Further details on the LCS characteristics and dynamics can be found in references \cite{Bao2019Laser,Rowley2022Self}.  Under typical operating conditions, we obtained a soliton output power of about 5 mW. The EDFA 980 nm pump for single soliton operation is typically requires 700 mW and is driven with 1 A at 1.8 V. To improve the mm-wave emission performance, we added an additional spectral shaping and amplification stage. Specifically, the output from the resonator's through-port was filtered through a bandpass filter centred at 1540 nm with a full width at half maximum of 12 nm, designed to suppress a strong unwanted mode arising from the erbium gain spectrum at 1540 nm. This filtering significantly improved SNR in subsequent THz-TDS  measurements. While the filtered output already provided sufficient power for basic THz generation and detection, a second fibre amplifier was included to increase flexibility, signal dynamics, and overall SNR, boosting the average optical power to approximately 120 mW for a single-soliton state. It is important to note that this power level varies substantially with the specific soliton configuration (e.g. the number of solitons in the state).
The filtered and amplified pulse train was subsequently directed into a conventional fibre-coupled THz-TDS setup. Here, a beam splitter separated the pulses into distinct pump and probe paths, with an optical delay stage controlling their relative timing. Both paths were fibre-coupled into commercial fibre-coupled photoconductive antennas (Menlo TERA15-TX-FC emitter and TERA15-RX-FC receiver). The emitted THz pulses propagated through free space and were collimated and focused using a pair of Teflon THz lenses. 
A photoconductive antenna generates THz transients by illuminating a biased semiconductor gap with a picosecond optical pulse, generating a transient photocurrent that radiates a THz pulse. In a common model, in the far-field the THz electric field is proportional to the time-derivative of the photocurrent: 
\begin{equation}
E(t) \propto \frac{d{J(t)}}{d{t}}. \label{eq3}
\end{equation}
The photocurrent $J(t)$ is roughly proportional to the photoexcited carrier density $n(t)$  a simple rate equation $(dn/dt  +n/\tau_e ) \propto I(t)$ describes the carrier dynamics. $I(t)$ is the optical intensity envelope and $\tau_e$ is the carrier lifetime. When  $\tau_e$ smaller than the typical minimum pump-pulse duration, we can assume a fast-recombination time. Considering a typical microcomb pulse with of $\Delta t=1/2$ ps, this is the typical regime in antennae specified for pumping at $\lambda=1550$ nm , typically based on Lt-InGaAs semiconductor substrates \cite{Xu202111}. Carriers are trapped or recombine quickly, preventing any delayed response, hence $J(t)\approx Const \times I(t)$, i.e the emitted field approximately follows the time-derivative of the optical intensity envelope in Eq.(1). The detected field in the photoconductive sampling in fast recombination regime is then the convolution of the THz field and the envelope of the sampling pulses, as indicated in Eq. (2). It is worth highlighting that our experimental configuration, that uses commercial photoconductive antennas typically specified for optical pumping repetition rates around 100 MHz, is notably sub-optimal for our $\sim$ 50 GHz microcomb-driven pulse train. Firstly, by constraining the average optical power to remain within the antenna's specified damage threshold, the energy per individual pulse (and thus the field amplitude per pulse) becomes significantly lower than what the antenna specifications. While partially compensated by the increased repetition rate, a typical side effect of the much shorter pulse period is the incomplete depletion of photo-generated carriers that establishes a higher conductivity background. This diminishes the achievable conductivity contrast, reducing the current gradient and thereby weakening the generated field emission. More critically, the electrical supply design of standard commercial antennas is commonly optimised for replenishing charges at the photoconductor electrodes within timescales compatible, with repetition rates of about 100 MHz. At our operational repetition rates—approximately 500 times higher—this replenishment time constant severely constrains the peak photocurrent achievable per pulse. This aspect severely bounds the antenna efficiency. Remarkably our results suggest that substantial improvements, possibly spanning several orders of magnitude, may be achievable through antenna and electrical supply optimisation specifically tailored to microcomb design.
\subsection{Microcomb Stabilisation Setup }
To stabilise the system, active control of the cavity length was implemented using a piezoelectric fibre stretcher. This stretcher comprises of a 62 cm-long segment of the Erbium-doped fibre amplifier wound around a 6 cm diameter piezoelectric cylinder. A proportional-integral-derivative (PID) control loop (mFALC 110) regulated the voltage applied to the cylinder, thereby continuously adjusting its radius to tune the overall cavity length.
The scheme for locking the soliton repetition rate is shown in Extended Figure A4. A portion of the resonator's through-port output was diverted using a beam splitter and directed into an electro-optic modulator (EOM). The EOM was operated in a high-drive regime at 8.168 GHz to generate sidebands on each soliton comb line. This drive frequency was precisely tuned such that the heterodyne beatnote between the outermost sidebands of adjacent $\sim$ 50 GHz-spaced comb lines was approximately 20 MHz. The modulated light was then directed to two photodiodes, and the output from each was passed through a 20–21 MHz electrical bandpass filter. One filtered signal was monitored with a frequency counter, while the other served as the process variable for the PID controller, which used a 20.5 MHz signal from a Sigilent generator as its reference setpoint. The resulting error signal from the PID was amplified to drive the piezoelectric fibre stretcher, thereby closing the feedback loop and locking the repetition rate beatnote to the 20.5 MHz reference.
It is important to note that both the frequency counter and the signal generator were synchronised to an external GPS-disciplined reference to minimise long-term noise contributions from the reference source in the locked measurements.
The overall duration of our measurements was ultimately limited by the long-term stability of this locking system. Over time, slow thermal drifts and environmental fluctuations caused the repetition rate to deviate beyond the dynamic range of the piezoelectric stretcher. When this compensation limit was exceeded, the feedback loop could no longer maintain its lock, preventing longer data acquisition times.

\backmatter

\bmhead{Data availability}
The data that support the finding of this study are openly available at doi: 10.17028/rd.lboro.30272239. Additional datasets generated during the current study are available from the corresponding author on reasonable request. 

\bmhead{Code availability}
The codes used during the current study are available from the corresponding author on reasonable request.

\bmhead{Contributions}
L.P., M.R., A.P., and MP. developed the original research idea. L.P led the microcomb to THz conversion experiments, A.C. led metrological microcomb experiments. A.C., A.R.C., V.C., contributed to the microcomb to THz conversion experiments. L.P., F.G., M.R., N.P., D.D., contributed to the metrological microcomb experiments. L.P., L.O., J.S.T.G., A.P. and M.P., analysed the experimental data. B.E.L. and S.T.C. designed and fabricated the integrated devices. L.P., L.O., M.P., and A.P. drafted the main paper, and all the authors contributed to the writing of the manuscript. A.P. and M.P. supervised the research.	

\bmhead{Funding}
 This project received funding from the European Research Council (ERC) under the European Union’s Horizon 2020 Research and Innovation Programme Grant No. 725046. The authors acknowledge financial support from the (UK) Engineering and Physical Sciences Research Council (EPSRC), Grant Nos. EP/Z533178/1, EP/Y004701/1 and EP/X012689/1,   the Leverhulme Trust (Research Project Grant number RPG-2022-090, Early Career Fellowship ECF-2020-537, Early Career Fellowship ECF-2022-710, Early Career Fellowship ECF-2023-315, Early Career Fellowship ECF-2024-529), Loughborough University’s Vice-Chancellor Independent Research Fellowship, and DEVCOM US Army Research Office, Grant agreement W911NF2310313.

\bibliography{sn-bibliography}
\newpage
\begin{appendices}

\section{Extended Figures}\label{secE1}

\begin{figure}[h]
\centering
\includegraphics[width=0.9\textwidth]{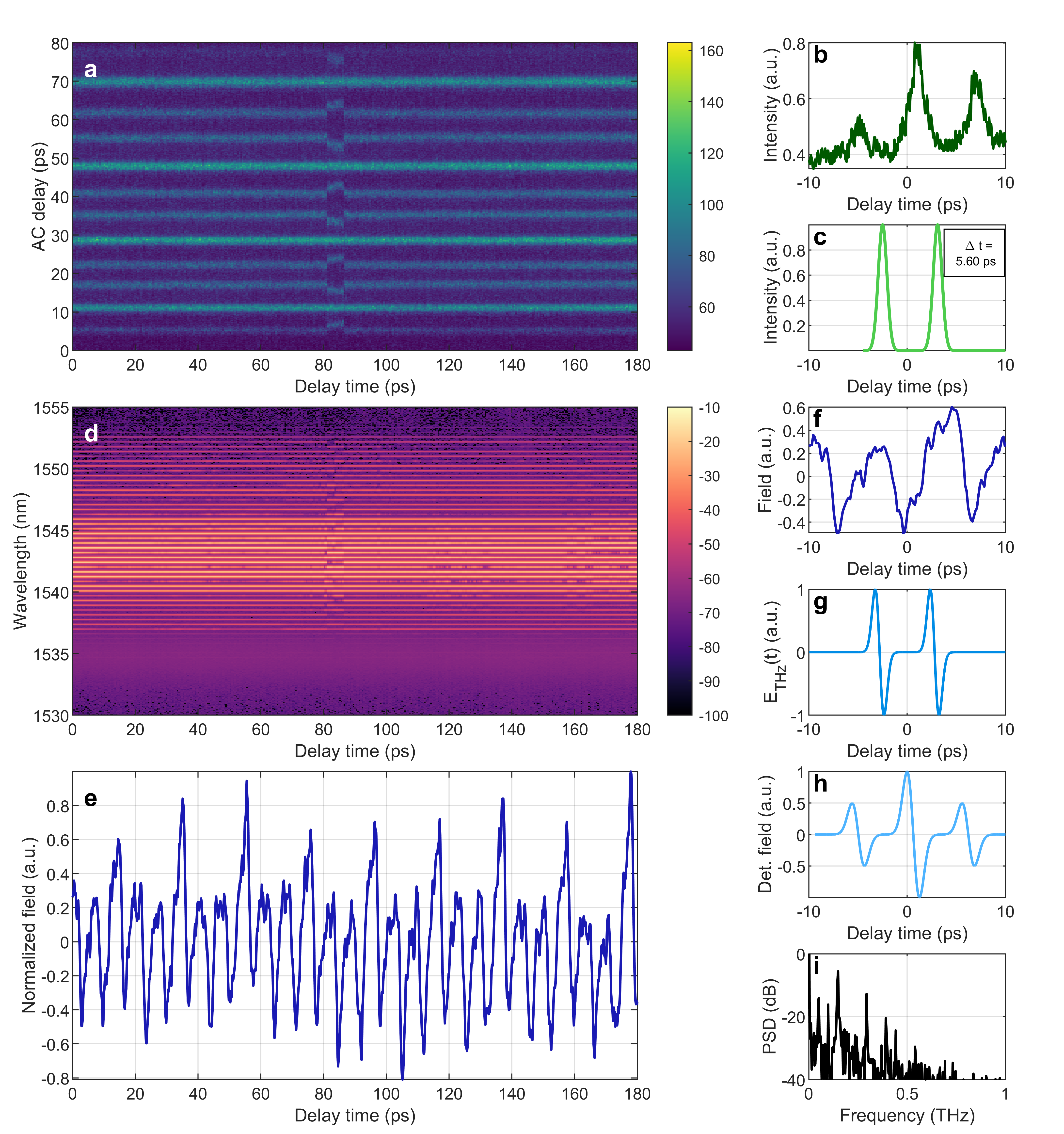}
\caption{\textbf{Alternate case for two solitons long term robustness} \textbf{a} a False-colour map of the autocorrelation traces in time and \textbf{b} A typical autocorrelation trace. \textbf{c} Reconstructed double soliton state with periodicity 20 ps and soliton spacing 4.46 ps extracted from the autocorrelation trace. \textbf{d} False-colour map of the optical spectrum analyser traces in time.
\textbf{e} Experimental terahertz time-domain spectroscopy trace. \textbf{f} Experimental TDS trace for a single period. \textbf{g} Expected generated waveform converting with Eq(1) the optical pulses in c.  \textbf{h}   Reconstructed TDS trace using Eq. (2) with the generated waveform in  g and the optical pulses in c. \textbf{i} Experimental THz spectrum.}\label{Efig1}
\end{figure}

\begin{figure}[h]
\centering
\includegraphics[width=0.9\textwidth]{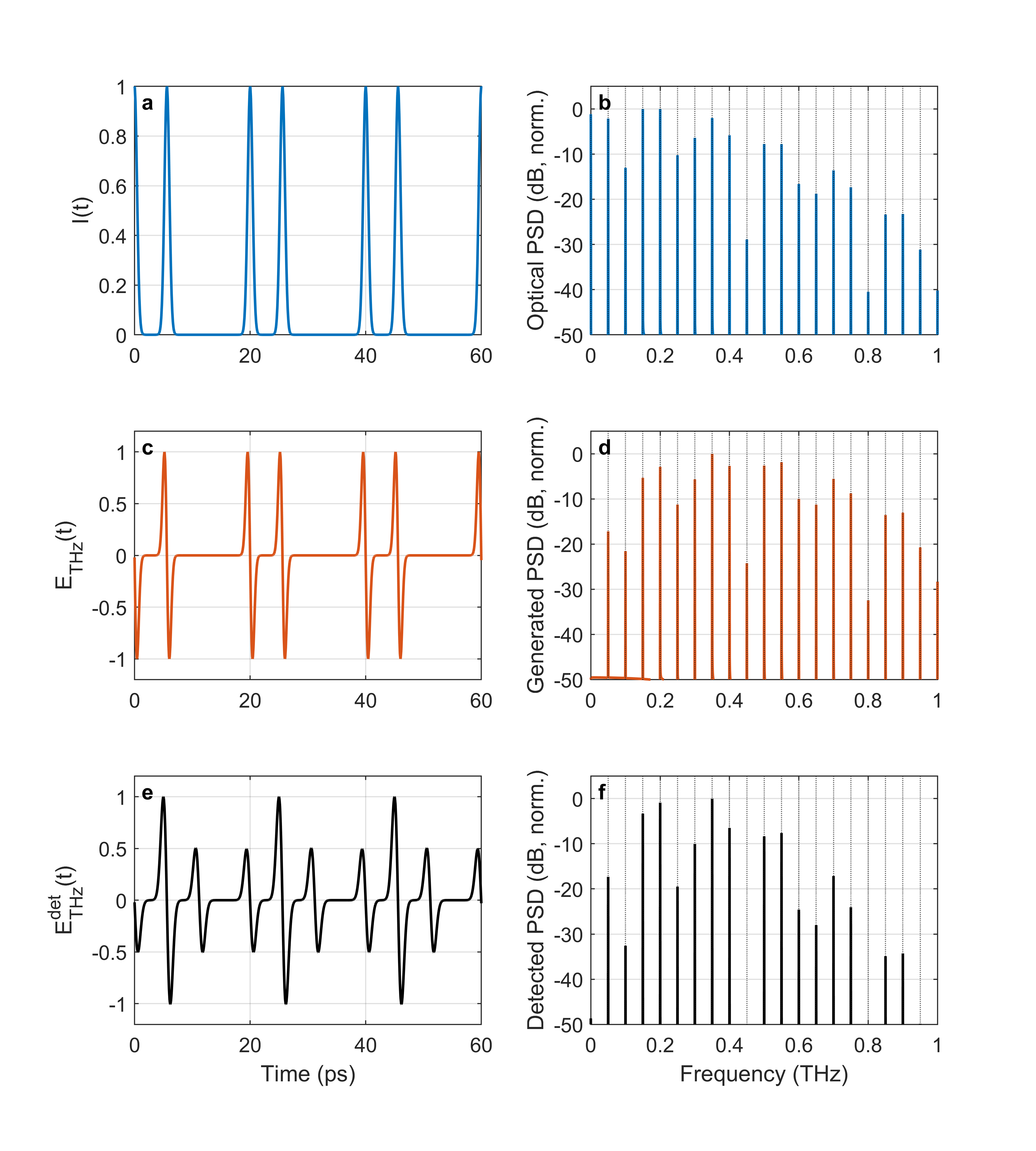}
\caption{\textbf{Single-case model of spectral control with a fixed soliton-pulse delay,} showing the effect of the optical–THz convolution in TDS reconstruction. \textbf{a} Normalised optical intensity over a 60 ps window.  \textbf{b} One-sided power spectral density of the optical pulse train. \textbf{c} Normalised generated THz waveform over a 60 ps window. \textbf{d} Power spectral density of the generated THz field. 
\textbf{e} Normalised detected THz waveform over a 60 ps window, obtained from the convolution of the optical intensity in a with the generated THz field in c. \textbf{f} Power spectral density of the detected TDS THz field, obtained from the multiplication of the optical pulse and the generated THz field. This convolution effectively reduces the visibility of some generated frequencies.}\label{Efig2}
\end{figure}

\begin{figure}[h]
\centering
\includegraphics[width=0.9\textwidth]{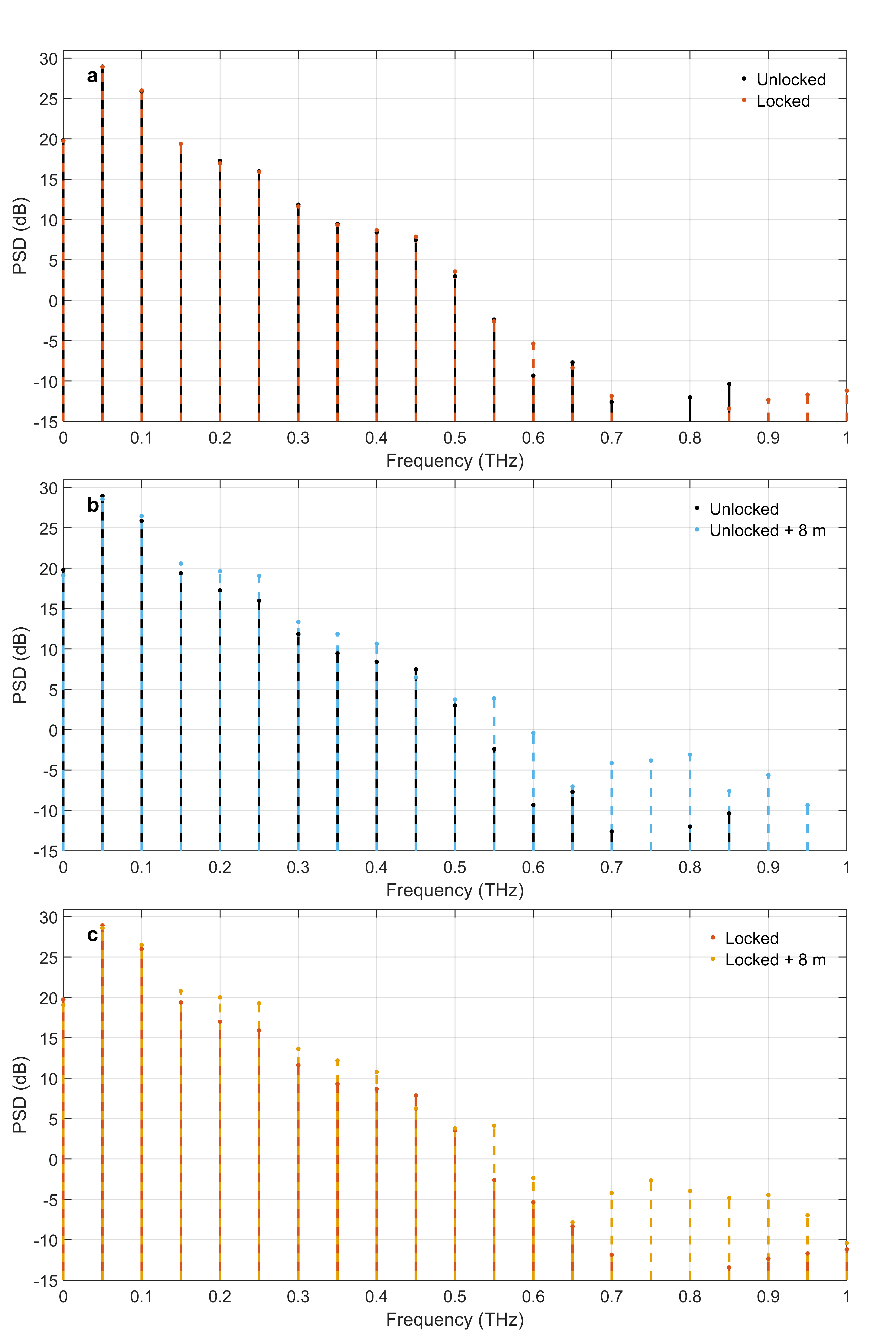}
\caption{\textbf{Comparison of millimetre‑wave spectra under different locking conditions} Here are shown the spectra of traces shown in Fig. d and e. \textbf{a} Unlocked (black) versus locked (vermillion) spectra. \textbf{b} Unlocked (black) versus unlocked with an added 8 m delay line (sky blue). \textbf{c} Locked (vermillion) versus locked + 8 m (orange). Spectra are plotted from 0–1 THz with vertical bar “stems” every 0.05 THz; dots mark the bar tips. No statistically meaningful deviation is observed after inserting the 8 m delay: residual amplitude changes are consistent with routine coupling‑power drift.}\label{Efig3}
\end{figure}

\begin{figure}[h]
\centering
\includegraphics[width=0.9\textwidth]{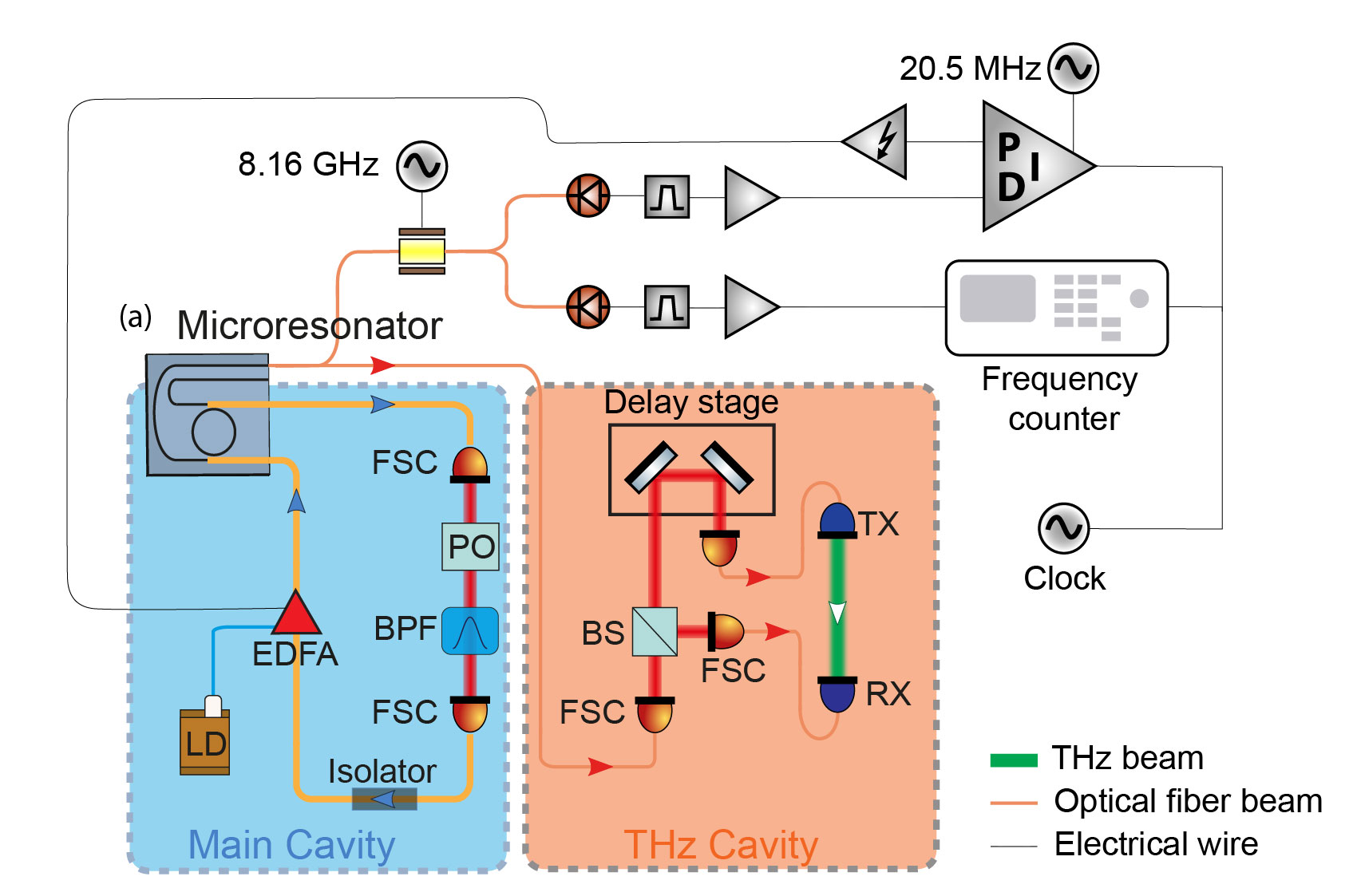}
\caption{\textbf{Microcomb‑driven THz setup and repetition‑rate lock} A laser‑cavity soliton is generated in a 48.9 GHz‑FSR silica micro‑ring that closes the main fibre cavity (blue). The cavity contains an EDFA, isolator, polarisation optics and a 10 nm band‑pass filter. Light from the through port is further filtered (12 nm Gaussian at 1540 nm) and optionally re‑amplified, then sent to a fibre‑coupled THz‑TDS stage (orange) where a beam splitter and delay line feed transmitter and receiver photoconductive antennas (TX/RX). To stabilise the repetition rate, a tap of the output is phase‑modulated at 8.168 GHz; the $\sim$ 20 MHz beat between adjacent comb lines is detected on photodiodes, filtered (20–21 MHz), and compared to a 20.5 MHz reference in a PID loop that drives a piezo fibre stretcher. All RF instruments share a GPS‑disciplined clock. Colours: green—THz beam; orange—optical fibre; black—electrical. Abbreviations: LD, laser diode; EDFA, Er/Yb‑doped fibre amplifier; BPF, band‑pass filter; PO, polarisation optics; BS, beam splitter; FSC, fibre splitter/combiner; PD, photodiode; PID, controller.}\label{Efig4}
\end{figure}




\end{appendices}



\end{document}